\documentclass[epj,twocolumn]{webofc}
\usepackage[varg]{txfonts}   
\usepackage{color}
\usepackage{tcolorbox}
\usepackage{xcolor}

\usepackage{physics}
\usepackage{slashed}
\usepackage{rotating}

\xdefinecolor{vscuro}{RGB}{30,90,0}

\newcommand{\Eqn}[1]{Eq.~(\ref{#1})}

\newcommand{\Br}{\text{Br}}

\woctitle{Flavour changing and conserving processes}
\begin{document}
\title{Effective-field theories for charged lepton flavour violation}
%
%

\author{Giovanni Marco Pruna\inst{1}}

\institute{INFN, Laboratori Nazionali di Frascati,
Via E. Fermi 40, 00044 Frascati (Rome), Italy}

\abstract{%
  These proceedings review the status of present and future bounds on muonic lepton flavour violating transitions in the context of an effective-field theory defined below the electroweak scale. A specific focus is set on the phenomenology of $\mu\to e\gamma$, $\mu \to 3e$ transitions and coherent $\mu\to e$ nuclear conversion in the light of current and future experiments. Once the experimental limits are recast into bounds at higher scales, it is shown that the interplay between the various experiments is crucial to cover all corners of the parameter space.
}
\maketitle
\section{Introduction}
\label{intro}
Lepton flavour violation (LFV) is strongly suppressed in the framework of the Standard Model (SM) of particle physics. Any observation of such phenomenon in the charged lepton sector would indicate unambiguously the presence of new physics (NP). Therefore, various experimental plans have taken place worldwide to confirm the vanishing prediction of the SM, and investigations are scheduled for the future to explore deeper regions of the NP parameter space.

Remarkable limits have been established with complementary tests of muonic LFV processes by the MEG~\cite{Adam:2013mnn,TheMEG:2016wtm} and SINDRUM~\cite{Bellgardt:1987du,Bertl:2006up} collaborations:
\begin{eqnarray}
	&&{\rm Br}\left( \mu^+ \to e^+ \gamma \right)\leq4.2 \times10^{-13}\,,\\
	&&{\rm Br}\left(\mu^+\to e^+e^-e^+ \right)\leq1.0 \times10^{-12}\,,\\
	&&{\rm Br}_{\mu \to e}^{\rm Au}  \equiv
    \frac{\Gamma(\mu^-\,{\rm Au}\to e^-\,{\rm Au})}{\Gamma^{\rm capt}_{{\rm Au}}}
     \leq 7 \times 10^{-13}\,.
\end{eqnarray}

Various experiments are already planned to improve these values by orders of magnitude: the MEG~II upgrade~\cite{Baldini:2013ke} with an expected sensitivity of ${\rm Br}(\mu \to e \gamma)\sim 5\times 10^{-14}$, the Mu3e experiment~\cite{Blondel:2013ia} with an improvement up to four orders of magnitude with respect to SINDRUM, and Mu2e at FNAL and COMET at J-PARC~\cite{Carey:2008zz,Kutschke:2011ux,Cui:2009zz} aiming to improve the sensitivity by four orders of magnitude compared with SINDRUM~II.

Consequently, these experimental efforts have to be supported both by an accurate theoretical interpretation of possible signals (or absence of signals) in terms of viable NP parameter space and a precise determination of the fundamental backgrounds\footnote{Recent progress has been made in the precise estimation of the $\mu\to e\gamma$~\cite{Fael:2015gua,Pruna:2017upz} and $\mu\to 3e$~\cite{Fael:2016yle,Pruna:2016spf} fundamental background.}.

These proceedings explore the possibility to give a model-independent phenomenological interpretation of muonic LFV transitions by adopting an effective-field-theory (EFT) description of NP interactions.

The EFT approach applied to LFV transitions has a long tradition: in the context of neutrino oscillations the first papers were published decades ago~\cite{Petcov:1976ff,Minkowski:1977sc}, while the first complete dimension-six parameterisation at low energy for charged LFV appeared in~\cite{Kuno:1999jp}. On the other hand, the first systematic treatments of charged LFV in the context of SM EFT~\cite{Buchmuller:1985jz,Grzadkowski:2010es} were published only a couple of years ago~\cite{Crivellin:2013hpa,Jenkins:2013zja,Jenkins:2013wua,Alonso:2013hga,Pruna:2014asa,Pruna:2015jhf}.

This note summarises the main results obtained in~\cite{Crivellin:2016ebg,Crivellin:2017rmk}, where the EFT parameterisation introduced by~\cite{Kuno:1999jp} was adopted to recast the current and future experimental limits on muonic LFV transitions in terms of bounds on the NP parameter space at the electroweak (EW) energy scale by exploiting a systematic renormalisation-group-equation (RGE) analysis.

\section{Parameterisation}
An effective Lagrangian for the $\mu\to e$ transitions valid below some scale $\Lambda$ with $m_{W} \ge \Lambda \gg m_b$ is considered. It consists of all the operators invariant under $U(1)_{\rm QED} \times SU(3)_{\rm QCD}$ and contains all the SM fermion fields (except for the top quark) and the QED and QCD gauge fields:
\begin{align}
&\mathcal{L}_{\rm eff}=\mathcal{L}_{\rm QED} + \mathcal{L}_{\rm QCD} 
\nonumber \\
&+\frac{1}{\Lambda^2}\bigg\{C_L^DO_L^D \nonumber \\
&+ \sum\limits_{f = q,\ell } {\left( 
      {C_{ff}^{V\;LL}O_{ff}^{V\;LL} + C_{ff}^{V\;LR}O_{ff}^{V\;LR} 
     + C_{ff}^{S\;LL}O_{ff}^{S\;LL}} \right)}\nonumber \\
& 
+ \sum\limits_{h = q,\tau } {\left( 
       {C_{hh}^{T\;LL}O_{hh}^{T\;LL} + C_{hh}^{S\;LR\;}O_{hh}^{S\;LR\;}} \right)}+
 L \leftrightarrow R\bigg\},
\label{Leff}
\end{align}
plus the Hermitian conjugate components wherever required, and the explicit form of the operators given with obvious notation by
\begin{align}
\label{eq:magnetic}
O_L^{D} &= e\, m_\mu\left( \bar e{\sigma ^{\mu \nu }}{P_L}\mu\right) {F_{\mu \nu }},
\\
O_{ff}^{V\;LL} &= \left(\bar e{\gamma ^\mu }{P_L}\mu\right) 
\left( \bar f{\gamma _\mu }{P_L}f\right),
\\
O_{ff}^{V\;LR} &= \left(\bar e{\gamma ^\mu }{P_L}\mu\right) 
\left( \bar f{\gamma _\mu }{P_R}f\right),
\\
O_{ff}^{S\;LL} &= \left(\bar e{P_L}\mu\right) \left( \bar f{P_L}f\right),
\\
O_{hh}^{S\;LR} &= \left(\bar e{P_L}\mu\right) \left( \bar h{P_R}h\right),
\\
O_{hh}^{T\;LL} &= \left(\bar e{\sigma _{\mu \nu }}{P_L}\mu\right) 
\left( \bar h{\sigma ^{\mu \nu }}{P_L}h\right),
\label{Ogg}
\end{align}
with $P_{L/R}=\left(\mathbb{I}\mp \gamma^5\right)/2$. In the equations above, $f$ represents any fermion below the scale $m_W$, and $h\in\{u,d,c,s,b,\tau\}$.

In the scenario where NP physics is realised at a scale $\Lambda < m_W$, NP gives rise to the interactions described in $\mathcal{L}_{\rm eff}$.
If BSM physics is beyond the EW scale, $SU(2)$-invariant higher-dimensional operators are generated in the SMEFT. Then, the higher-dimensional operators in $\mathcal{L}_{\rm eff}$ stem from the matching of the SMEFT to our theory, as performed at the tree level in~\cite{Davidson:2016edt}.

\section{Observables}
Given that the expressions for the rates of the processes $\mu^+ \to e^+\gamma$, $\mu^+\to e^+ e^- e^+$, and coherent muon-to-electron conversion in muonic atoms $\mu^- N\to e^- N$ will be exploited in the forthcoming section, we present them in terms of the effective coefficients (in the limit $m_e\ll m_\mu$).

For $\mu^+\to e^+\gamma$, at the tree level, the branching ratio is
\begin{align}
\label{muegBR}
{\rm{Br}}\left( {\mu  \to e\gamma } \right)  =  
\frac{\alpha_e m_\mu ^5}{{\Lambda^4 \Gamma _\mu }}\left( 
{{{\left| {C^{D}_L} \right|}^2} 
+ {{\left| {C^{D}_R} \right|}^2}} \right)\,,
\end{align}
where $\Gamma _\mu$ is the width of the muon.

The branching ratio of $\mu\to 3e$ expressed in terms of effective coefficients is
\begin{align}
&{\rm Br}(\mu  \to 3e) =\nonumber \\
&\frac{\alpha_e^2 m_\mu^5 }{12 \pi \Lambda^4 \Gamma _\mu}
\left(\left|C^{D}_{L}\right|^2+\left|C^{D}_{R}\right|^2\right)
\left(8 \log\left[\frac{m_\mu}{m_e}\right]-11\right)\nonumber
\\ 
&+\frac{m_\mu^5}{3  (16\pi)^3 \Lambda^4 \Gamma _\mu}
\bigg(\, \left|C_{ee}^{S\;LL}\right|^2 + 
16 \left|C_{ee}^{V\;LL}\right|^2 + 8 \left|C_{ee}^{V\;LR}\right|^2  \nonumber\\
&
+\
\left|C^{S\;RR}_{ee}\right|^2 +
16 \left|C_{ee}^{V\;RR}\right|^2 + 8 \left|C_{ee}^{V\;RL}\right|^2\bigg)+ X_\gamma \, ,
\end{align}
where the interference term with the dipole operator is given by
\begin{align}
X_\gamma ^{}{\rm{ }} =
-\frac{\alpha_e m_\mu^5 }{3 (4\pi)^2 \Lambda^4 \Gamma _\mu}
\left(\Re[C^{D}_{L}\left(C_{ee}^{V\;RL}+2C_{ee}^{V\;RR}\right)^*]+\right.\nonumber \\
\left.\Re[C^{D}_{R}
  \left(2C_{ee}^{V\;LL}+C_{ee}^{V\;LR}\right)^*]\right)\, .
\end{align}

Once relativistic and finite nuclear size effects are taken into account for heavy nuclei~\cite{Shanker:1979ap,Czarnecki:1998iz, Kitano:2002mt}, the transition amplitudes for the coherent $\mu\to e$ transition in nuclei exhibit different sensitivities with respect to the atomic number involved. Consequently, different target atoms provide different limits on the coefficients of the involved class of operators.
The SINDRUM collaboration has presented limits for gold, titanium and lead~\cite{Bertl:2006up,Dohmen:1993mp, Honecker:1996zf}, but the upcoming experiments mostly concentrate on aluminium.

For this process, the Lagrangian $\mathcal{L}_{\rm eff}$ as given in \Eqn{Leff} is not directly applicable. Instead, a Lagrangian at the nucleon level containing proton and neutron fields is required. This Lagrangian is obtained in two steps: integrating out heavy quarks and matching at a scale of $\mu_n=1$~GeV to an effective nuclear Lagrangian. Following~\cite{Cirigliano:2009bz}, the transition rate $\Gamma_{\mu\to e}^N = \Gamma(\mu^- N \to e^- N)$ is
\begin{align}
\Gamma_{\mu\to e}^N &= \frac{m_{\mu}^{5}}{4\Lambda^4}
\left| 4\left(
G_F m_\mu m_p \tilde{C}_{(p)}^{SL}  S^{(p)}_N
+   \tilde{C}_{(p)}^{VR} \; V^{(p)}_N
+ p \to n \right) \right.\nonumber \\
&\left.\, \, \, \, \, \, \, \qquad+e\,  C^{D}_{L} \; D_N 
 \right|^2 +L\leftrightarrow R,
\label{Gconv}
\end{align}
where $p$ and $n$ denote the proton and the neutron, respectively. The
effective couplings in \Eqn{Gconv} can be expressed in terms of our
Wilson coefficients as
\begin{align}
\tilde{C}_{(p/n)}^{VR} &= \sum_{q=u,d,s} 
\left(C_{qq}^{V\;RL}+C_{qq}^{V\;RR}\right) \; f^{(q)}_{Vp/n} \, , 
\label{tildeCVR} \\
\tilde{C}_{(p/n)}^{SL} &=  \sum_{q=u,d,s} 
\frac{\left(C_{qq}^{S\;LL}+C_{qq}^{S\;LR}\right)}{m_\mu m_q G_F} 
\; f^{(q)}_{Sp/n}  \nonumber\\
& - \frac{1}{12 \pi}
\sum_{q=c,b} \frac{C_{qq}^{S\;LL}+C_{qq}^{S\;LR}}{G_F\, m_\mu m_q} \;f_{Gp/n}\, ,
\label{tildeCSL}
\end{align}
with analogous relations for $L\leftrightarrow R$.

The nucleon form factors can be recast from~\cite{Hoferichter:2015dsa,Junnarkar:2013ac}. A next-to-leading-order computation of the nuclear Wilson coefficients appeared recently in the literature~\cite{Bartolotta:2017mff}; however, the inclusion of this result does not change any qualitative conclusion of the present study.

The branching ratio used in the following section is defined as the transition rate in~\Eqn{Gconv}, divided by the capture rate of the considered atom. For the latter, the values taken from~\cite{Suzuki:1987jf} are adopted.

\section{Connecting different energy scales}
The operators in $\mathcal{L}_{\rm eff}$ carry the phenomenological information obtained from the experimental constraints. At the tree level, however, a direct interpretation would result in bounds on the Wilson coefficients evaluated at the experimental scales, \emph{i.e.} $C(\lambda=m_\mu)$ or $C(\lambda=\mu_n)$.

A more meaningful procedure consists of extracting limits on the Wilson coefficients at scales that are different/higher than the phenomenological scale. If the NP is not far above the EW scale, the interconnection among scales is realised by the QED and QCD RGE of the effective operators, with the coefficients evolved between the experimental scale and the NP scale, \emph{i.e.} $m_{\mu},\mu_N\leq\Lambda \leq m_W$.

Under the RGE, the various operators in $\mathcal{L}_{\rm eff}$ mix among each other. To encode the leading mixing effects, at least the one-loop anomalous dimensions for all the operators must be considered. Moreover, the dipole operator plays a prominent role in all $\mu\to e$ transitions, and two-loop effects of direct mixing into $C^D_L$ and $C^D_R$ that are at the leading order must be considered (also to ensure a regularisation-scheme-independent result~\cite{Misiak:1993es, Ciuchini:1993fk}).

Including only a subset of two-loop leading-order contributions, while phenomenologically useful, is not a self-consistent procedure and should not be understood as a replacement of a genuine two-loop precise calculation, but rather as a qualitative indication of leading-order effects.

The evolution of the full set of operators is described by the following equations, where
\begin{align}
\dot{C} \equiv (4\pi)\,  \mu \dv{}{\mu}\, C
\end{align}
is understood, and the remaining notation is clear.

In the BMHV scheme~\cite{tHooft:1972tcz,Breitenlohner:1977hr}, the coefficient of the dipole operator runs according to
\begin{align}\label{rgedipole}
\dot{C}^{D}_{L}&=
16\, \alpha_{e}\, Q_l^2 C^{D}_{L}
-\frac{Q_l}{(4\pi)}\frac{m_{e}}{m_{\mu}}C^{S\;LL}_{ee}
-\frac{Q_l}{(4\pi)} C^{S\;LL}_{\mu\mu} \nonumber\\
&+ \sum_{h} \frac{8 Q_h}{(4\pi)} \frac{m_{h}}{m_{\mu}} 
   N_{c,h} \, C^{T\;LL}_{hh} \,\Theta(\mu-m_h)\nonumber\\
&- \frac{\alpha_e Q_l^3}{(4\pi)^2} \left(
    \frac{116}{9} C_{ee}^{V\;RR} 
  + \frac{116}{9} C_{\mu\mu}^{V\;RR}\right.\nonumber \\
&\left.
  - \frac{122}{9} C_{\mu\mu}^{V\;RL}
  - \left(\frac{50}{9} + 8\, \frac{m_e}{m_\mu}\right) 
     C_{ee}^{V\;RL}\right) \nonumber\\
&- \sum_h \frac{\alpha_e}{(4\pi)^2}  \left(6 Q_h^2 Q_l
      + \frac{4 Q_h Q_l^2}{9}\right) N_{c,h} 
      \,  C_{hh}^{V\;RR} \,\Theta(\mu-m_h)\nonumber\\
&- \sum_h \frac{\alpha_e}{(4\pi)^2} \left(- 6 Q_h^2  Q_l
      + \frac{4 Q_h Q_l^2}{9}\right) N_{c,h} 
   \, C_{hh}^{V\;RL} \,\Theta(\mu-m_h)\nonumber\\
&- \sum_h \frac{\alpha_e}{(4\pi)^2} \, 4 Q_h^2 Q_l
   N_{c,h}\frac{m_{h}}{m_{\mu}} C^{S\;LR}_{hh}\,\Theta(\mu-m_h) \, .
\end{align}

The running of the whole set of vector operators is given by the
following two equations:
\begin{align}
\dot{C}^{V\;RR}_{ff}&=
\frac{4\, \alpha_{e}}{3}Q_f\left(
2 Q_l \sum_{\ell=e,\mu}C^{V\;RR}_{\ell\ell}+Q_l C^{V\;RR}_{\tau\tau}
+Q_l \sum_{l}C^{V\;RL}_{ll}\right.\nonumber\\
&\left.+N_c \sum_{q} Q_q\left(C^{V\;RR}_{qq}+C^{V\;RL}_{qq}\right)
+9 Q_l C^{V\;RR}_{ff}
\right),
\label{appVRR}
\end{align}
\begin{align}
\dot{C}^{V\;RL}_{ff}&=
\frac{4\, \alpha_{e}}{3}Q_f \left(
2 Q_l \sum_{\ell=e,\mu}C^{V\;RR}_{\ell\ell}+Q_l C^{V\;RR}_{\tau\tau}
+Q_l \sum_{l}C^{V\;RL}_{ll}\right.\nonumber\\
&\left.+N_c \sum_{q} Q_q \left(C^{V\;RR}_{qq}+C^{V\;RL}_{qq}\right)
-9 Q_l C^{V\;RL}_{ff}
\right).
\label{appVRL}
\end{align}

The running of the leptonic scalar and tensorial operators is
summarised by the following equations: 
\begin{align}
\dot{C}^{S\;LL}_{\ell\ell}&=
  12\,\alpha_{e}\,Q_l^2 C^{S\;LL}_{\ell\ell} \qquad \mbox{for\ }   \ell\in\{e,\mu\},
\\
\dot{C}^{S\;LL}_{\tau\tau}&=
-12\,\alpha_{e}\,Q_l^2 
\left(
C^{S\;LL}_{\tau\tau}+8C^{T\;LL}_{\tau\tau}
\right),
\\
\dot{C}^{S\;LR}_{\tau\tau}&=
-12\,\alpha_{e}\,Q_l^2 C^{S\;LR}_{\tau\tau},
\\
\dot{C}^{T\;LL}_{\tau\tau}&=
-2\,\alpha_{e}\,Q_l^2 
\left(
C^{S\;LL}_{\tau\tau}-2C^{T\;LL}_{\tau\tau}
\right).
\end{align}
The running of the scalar and tensorial quark operators is given by 
\begin{align}
\dot{C}^{S\;LL}_{qq}&=
\left(-6\,\alpha_{e}\left(Q_l^2+Q_q^2\right)
-6 C_F\,\alpha_s \right)C^{S\;LL}_{qq}\nonumber \\
&-96\,\alpha_{e}\,Q_l Q_q C^{T\;LL}_{qq},
\\
\dot{C}^{S\;LR}_{qq}&=
\left(-6\,\alpha_{e}\left(Q_l^2+Q_q^2\right)
-6 C_F\,\alpha_s\right)C^{S\;LR}_{qq},
\\
\dot{C}^{T\;LL}_{qq}&=
-2\,\alpha_{e}\,Q_l Q_qC^{S\;LL}_{qq}\nonumber \\
&+\left(2\,\alpha_{e}\left(Q_l^2+Q_q^2\right)+2C_F\,\alpha_s\right)C^{T\;LL}_{qq}.
\end{align}

\section{Phenomenological results}
The effective coefficients are generated by some underlying NP theory around the EW scale. Then, the RGEs can be exploited to evolve such coefficients from the high scale $m_W$ to the phenomenological scales $\mu_n$ and $m_\mu$. Therefore, the predicted rates are compared with the experimental limits. This procedure will shape the constraints of various Wilson coefficients at the NP scale.

RGE mixing can generate important effects at the phenomenological scale even for vanishing Wilson coefficients at the high scale. This allows one to place bounds on coefficients that would be unconstrained if loop effects are not taken into account.

Considering the experimental limits listed in Section~\ref{intro}, the final aim is to compare the exploratory power of current and future $\mu\to e\gamma$, $\mu\to 3e$ and $\mu\to e$ conversion experiments for specific effective interactions at the NP scale.

\begin{figure*}[!th]
\begin{center}
\includegraphics[width=\textwidth]{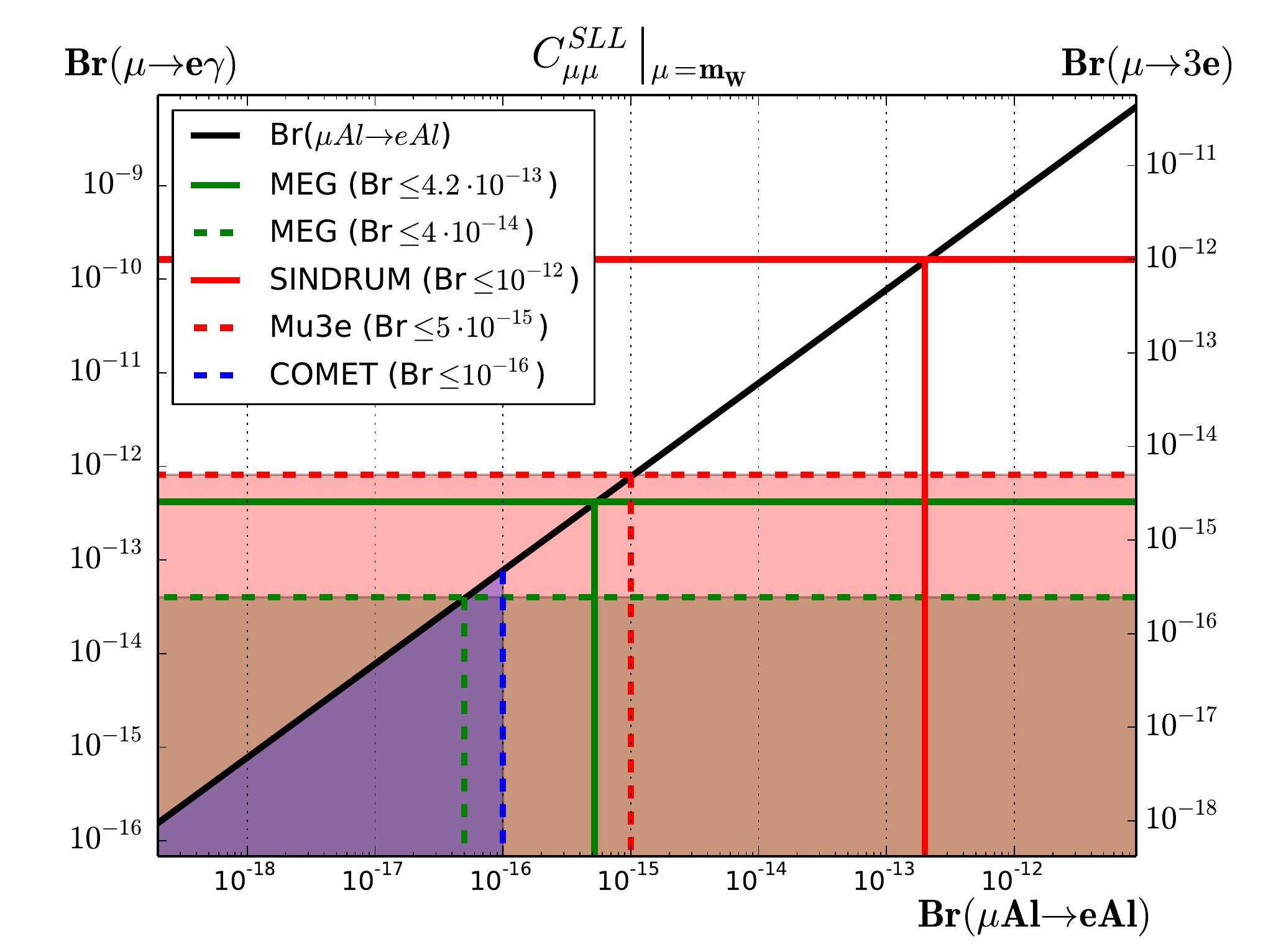}
\end{center}
\caption{\Br($\mu\to e\gamma$) (\Br($\mu\to 3e$)) plotted on the left
  (right) $y$-axis against \Br($\mu N\to eN$) for a fixed value of
  $C^{S\; LL}_{\mu\mu}$ given at the scale $\mu=m_W$. Current and future
  experimental limits are displayed.}
\label{figE}
\end{figure*}

First, we consider the coefficient for which the MEG experiment and future updates will deliver the best performances: $C^{S\; LL}_{\mu\mu}$. In Figure~\ref{figE}, the current and future branching ratio for $\mu\to e\gamma$ and $\mu\to 3e$ experiments are compared with future $\mu N\to e N$ prospects.

Here, the horizontal dashed-red line indicates that a small limit $\Br(\mu\to e\gamma)\lesssim 10^{-12}$ displays an equivalent exploratory power as the future Mu3e limit $\Br(\mu\to 3e)< 5\times 10^{-15}$. Instead, for muon conversion experiments (vertical dashed-red line), a limit of $\Br(\mu N\to eN)< 10^{-15}$ would be required to perform as good as $\mu\to e \gamma$ experiments. In other words, the future MEG~II experiment will place the strongest limit on $C^{S\,LL}_{\mu\mu}$ unless the COMET or Mu2e experiments could improve their expected limit to reach at least $\Br(\mu N\to eN)<5\times10^{-17}$. Curiously, one must notice that experiments exploring the dipole interaction at lower scales will deliver the best future performance in testing a four-fermion interaction defined at higher scales. This goes against one of the most deeply embedded prejudices in the theoretical and experimental LFV community.

Now, we considered scenarios where two Wilson coefficients are non-vanishing at the EW scale. For this scope, the allowed parameter space is plotted in light of current and future experimental limits for all three processes. For clarity's sake, they are displayed on a pseudo-logarithmic scale.

\begin{figure*}[!th]
\begin{center}
\includegraphics[width=\textwidth]{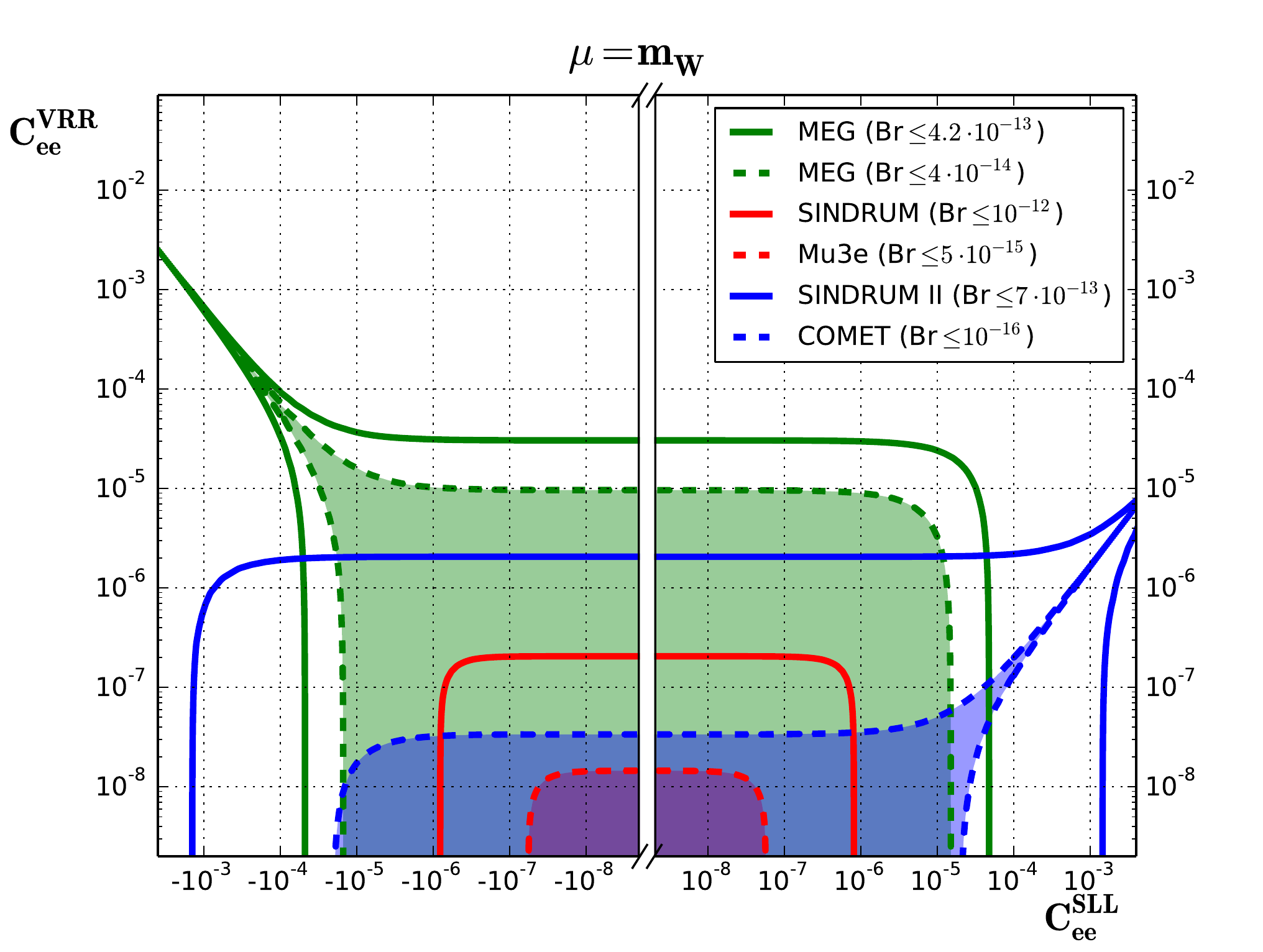} 
\end{center} 
\caption{Allowed regions in the $C^{S\,LL}_{ee}-C^{V\,RR}_{ee}$ plane
  from $\mu\to e\gamma$ (green), $\mu\to3e$ (red) and $\mu\to e$
  conversion (blue) for current (straight) and future (dashed)
  experimental limits.}
\label{figB}
\end{figure*}
In Figure~\ref{figB}, the allowed regions in the $C^{S\,LL}_{ee}-C^{V\,RR}_{ee}$ plane are plotted (with coefficients defined at the EW scale). Current (solid lines) and future (dashed lines) limits on $\mu\to e\gamma$ (green) and $\mu\to e$ conversion (blue) are less stringent than those coming from $\mu\to3e$ (red). This indicates that $\mu \to 3e$ experiments are and will be the most sensitive to these Wilson coefficients. Moreover, $\mu \to 3e$ experiments do not provide any blind region of the parameter space for this specific choice of coefficients. This occurs because these operators produce $\mu\to 3e$ already at the tree level, while they give rise to the other processes only via mixing effects.

\begin{figure*}[!th]
\begin{center}
\includegraphics[width=\textwidth]{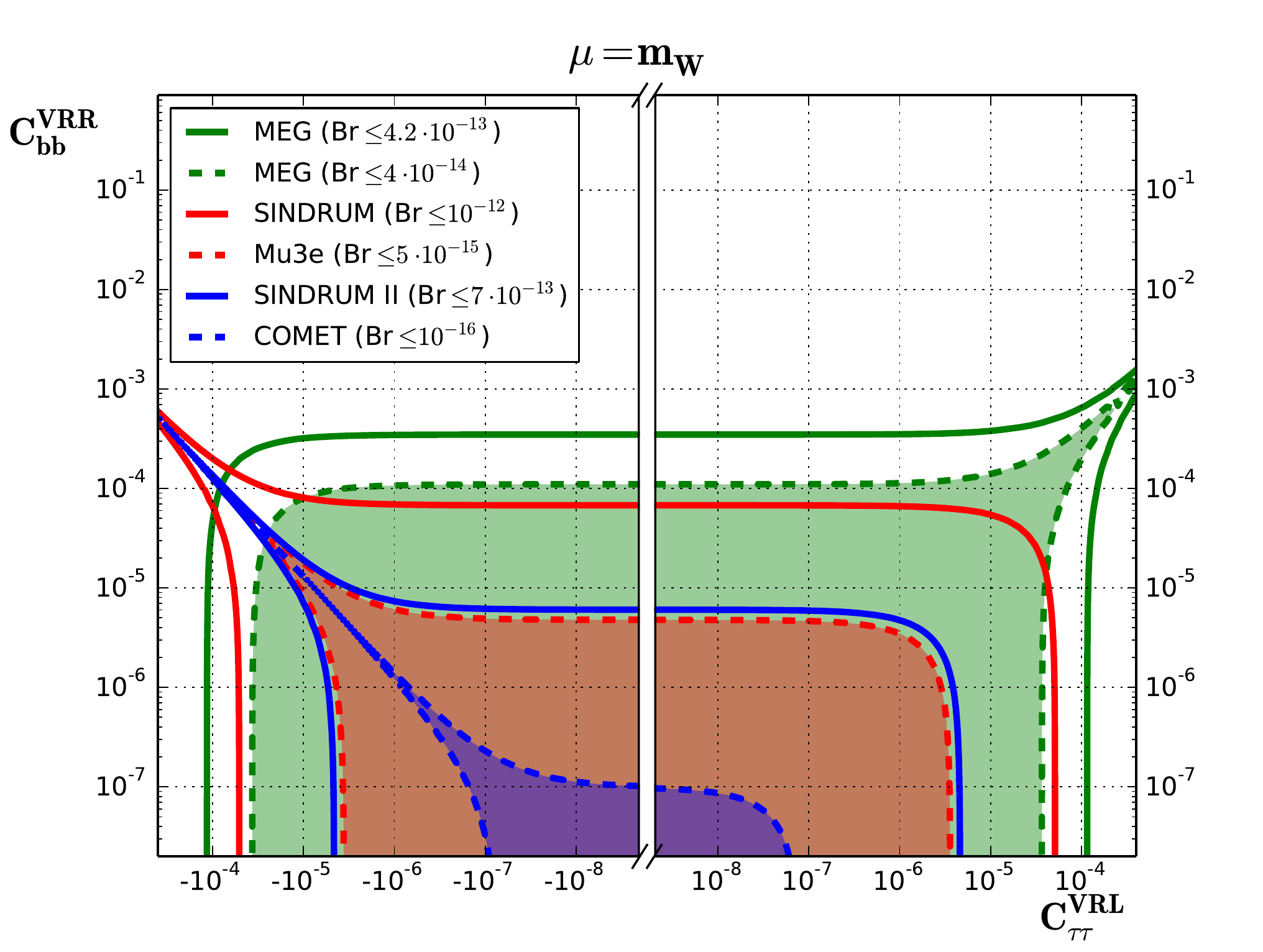}
\end{center}
\caption{Allowed regions in the
  $C^{V\,RL}_{\tau\tau}-C^{V\,RR}_{bb}$ plane from $\mu\to e\gamma$
  (green), $\mu\to 3e$ (red) and $\mu\to e$ conversion (blue) for
  current (straight) and future (dashed) experimental limits.}
\label{figA}
\end{figure*}

\begin{figure*}[!th]
\begin{center}
\includegraphics[width=\textwidth]{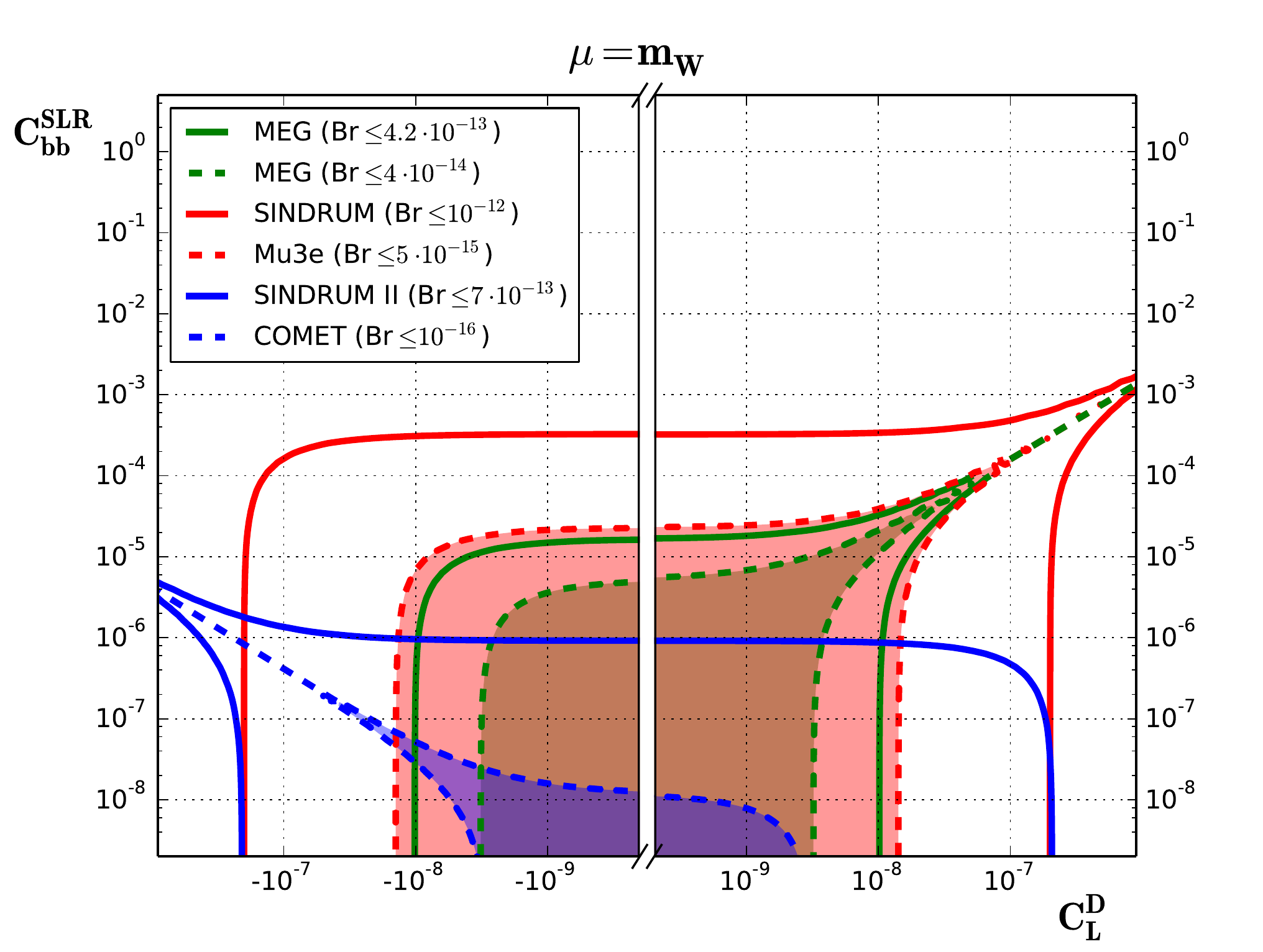}
\end{center}
\caption{Allowed regions in the $C^{D}_{L}-C^{V\,RR}_{ee}$ plane
  from $\mu\to e\gamma$ (green), $\mu\to 3e$ (red) and $\mu\to e$
  conversion (blue) for current (straight) and future (dashed)
  experimental limits.}
\label{figD}
\end{figure*}
Figure~\ref{figA} shows an analogous plot for the Wilson coefficients $C^{V\,RL}_{\tau\tau}$ and $C^{V\,RR}_{bb}$. Here, $\mu\to e$ conversion experiments display a superior capability to probe vectorial four-fermion operators generated at the phenomenological scale by mixing effects. Even the future Mu3e experiment will perform just around the current $\mu\to e$ conversion limit established by the SINDRUM~II collaboration more than a decade ago. This shows that future Mu2e and COMET experiments will represent the best opportunity to test four-fermion vectorial interactions containing \emph{any type of heavy fermion}, including $\tau$ leptons. Again, this result challenges the biased opinion that coherent LFV conversion in nuclei can test only contact interactions with valence quarks.

Furthermore, the plot displays an interesting complementarity among various experiments: assuming that the underlying theory produces a cancellation both in $\mu\to e$ conversion and $\mu\to 3e$, then $\mu\to e\gamma$ experiments will provide a complementary limit, ultimately circumscribing the allowed region of the parameter space.


Figure~\ref{figD} shows the allowed regions in the $C^{D}_{L}-C^{S\:LR}_{bb}$
plane. For this choice of coefficients, one may immediately notice that $\mu\to3e$ experiments are less constraining than the other two options. On the other hand, the COMET and Mu2e experiments will indeed set the best limits on each uncorrelated Wilson coefficient. However, there is a big portion of the parameter space where $\mu\to e$ conversion experiments are completely blind. To cover this region, results from MEG~II will be required.

\begin{figure*}[!th]
\begin{center}
\includegraphics[width=\textwidth]{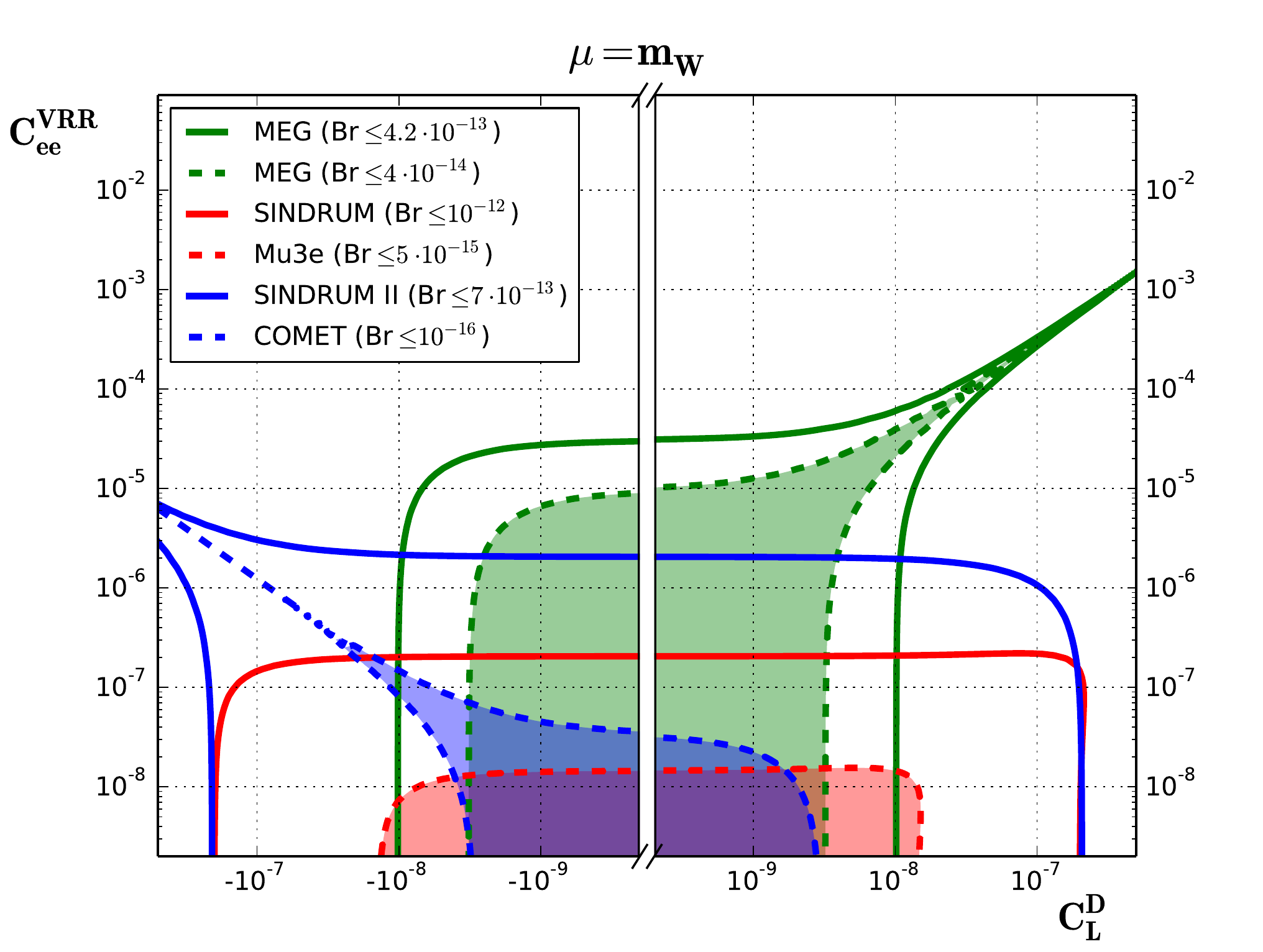}
\end{center}
\caption{Allowed regions in the $C^{D}_{L}-C^{V\,RR}_{ee}$ plane
  from $\mu\to e\gamma$ (green), $\mu\to 3e$ (red) and $\mu\to e$
  conversion (blue) for current (straight) and future (dashed)
  experimental limits.}
\label{figC}
\end{figure*}
With Figure~\ref{figC}, we conclude our overview by considering the coefficients $C^{D}_{L}$ and $C^{V\,RR}_{ee}$. In this case, it is well established that the current limit on the $C^{D}_{L}$ coefficient comes from the MEG experiment, and the future limit will be set by Mu2e and COMET. Focusing on the $C^{V\,RR}_{ee}$ coefficient, even with intuitive arguments one must realise that $\mu \to 3e$ experiments (SINDRUM in the past and Mu3e in the future) give the most significant limits. However, in the corners of the parameter space where potential cancellations might occur, an interesting interplay between the observables implies that all of the future experimental limits are useful to ensure that no blind spots in parameter space remain unexplored.

Of course, our choice of combinations for the free parameters is far from being exhaustive. However, the presented analysis carries two main messages:
\begin{itemize}
\item many opinions that are deeply entrenched in the LFV theoretical and experimental community are undermined by a systematic RGE treatment of effective coefficients;
\item the interplay between the various experiments is crucial to cover all corners of the NP parameter space, especially where cancellations can result in blind spots for one or even two specific experiments.
\end{itemize}

\section{Conclusion}

In these proceedings, an RGE-improved analysis of the three $\mu\to e$ processes $\mu\to e\gamma$, $\mu\to 3e$ and $\mu\to e$ conversion in nuclei in the context of EFT was briefly reviewed.

The complete set of dimension-six operators giving rise to point-like muonic LFV interactions invariant under $U(1)_{\rm QED} \times SU(3)_{\rm QCD}$ was introduced.

The relevant observables for muonic LFV transitions and the formulae for their transition rates were recollected.

The complete one-loop RGEs for an EFT with dimension-six operators invariant under $U(1)_{\rm QED} \times SU(3)_{\rm QCD}$ were presented. Furthermore, the leading two-loop QED effects for the mixing of vector operators into the dipole operators were included.

Then, the resulting bounds on the Wilson coefficients (given at the scale $m_W$) were calculated. Afterwards, some benchmark scenario was considered to stress both the complementarity of the three $\mu\to e$ processes and the inaccuracy of some opinions that are widely held in the LFV community. Concerning the former, the capability of covering regions of parameter space that would be blind spots for a single process was clearly pointed out. Concerning the latter, the potential of $\mu\to e\gamma$ experiments to explore scalar four-fermion interactions and the impact from coherent $\mu \to e$ conversion on vectorial four-fermion interactions with heavy fermions was clearly displayed.

The present analysis is far from being exhaustive; indeed, intriguing new ideas to explore the effective parameter space appeared recently in the literature~\cite{Koike:2010xr,Uesaka:2016vfy,Uesaka:2017yin,Cirigliano:2017azj,Davidson:2017nrp}, and their RGE-improved analysis should be included in the present treatment.

Finally, even if a systematic study was never performed in this direction, we claim that future developments of the $\mu e$-scattering (MUonE) experiment~\cite{Calame:2015fva,Abbiendi:2016xup} could increase our current knowledge of muon LFV transitions and a dedicated EFT analysis should be performed in a future study.

%
%
%
\bibliography{biblio}

\begin{thebibliography}{48}

\bibitem{Adam:2013mnn}
J.~Adam et~al. (MEG), Phys. Rev. Lett. \textbf{110}, 201801 (2013),
  \texttt{1303.0754}

\bibitem{TheMEG:2016wtm}
A.~M. Baldini et~al. (MEG), Eur. Phys. J. \textbf{C76}, 434 (2016),
  \texttt{1605.05081}

\bibitem{Bellgardt:1987du}
U.~Bellgardt et~al. (SINDRUM), Nucl. Phys. \textbf{B299}, 1 (1988)

\bibitem{Bertl:2006up}
W.~H. Bertl et~al. (SINDRUM II), Eur. Phys. J. \textbf{C47}, 337 (2006)

\bibitem{Baldini:2013ke}
A.~M. Baldini et~al. (2013), \texttt{1301.7225}

\bibitem{Blondel:2013ia}
A.~Blondel et~al. (2013), \texttt{1301.6113}

\bibitem{Carey:2008zz}
R.~M.~Carey et al. (Mu2e), FERMILAB-PROPOSAL-0973 (2008)

\bibitem{Kutschke:2011ux}
R.~K. Kutschke, \emph{{The Mu2e Experiment at Fermilab}}, in \emph{{Proceedings,
  31st International Conference on Physics in collisions (PIC 2011): Vancouver,
  Canada, August 28-September 1, 2011}} (2011), \texttt{1112.0242}

\bibitem{Cui:2009zz}
Y.~G.~Cui et al. (COMET), KEK-2009-10 (2009)

\bibitem{Fael:2015gua}
M.~Fael, L.~Mercolli, M.~Passera, JHEP \textbf{07}, 153 (2015),
  \texttt{1506.03416}

\bibitem{Pruna:2017upz}
G.~M. Pruna, A.~Signer, Y.~Ulrich, Phys. Lett. \textbf{B772}, 452 (2017),
  \texttt{1705.03782}

\bibitem{Fael:2016yle}
M.~Fael, C.~Greub, JHEP \textbf{01}, 084 (2017), \texttt{1611.03726}

\bibitem{Pruna:2016spf}
G.~M. Pruna, A.~Signer, Y.~Ulrich, Phys. Lett. \textbf{B765}, 280 (2017),
  \texttt{1611.03617}

\bibitem{Petcov:1976ff}
S.~T. Petcov, Sov. J. Nucl. Phys. \textbf{25}, 340 (1977), [Erratum: Yad.
  Fiz.25,1336(1977)]

\bibitem{Minkowski:1977sc}
P.~Minkowski, Phys. Lett. \textbf{67B}, 421 (1977)

\bibitem{Kuno:1999jp}
Y.~Kuno, Y.~Okada, Rev. Mod. Phys. \textbf{73}, 151 (2001),
  \texttt{hep-ph/9909265}

\bibitem{Buchmuller:1985jz}
W.~Buchmuller, D.~Wyler, Nucl. Phys. \textbf{B268}, 621 (1986)

\bibitem{Grzadkowski:2010es}
B.~Grzadkowski, M.~Iskrzynski, M.~Misiak, J.~Rosiek, JHEP \textbf{10}, 085
  (2010), \texttt{1008.4884}

\bibitem{Crivellin:2013hpa}
A.~Crivellin, S.~Najjari, J.~Rosiek, JHEP \textbf{04}, 167 (2014),
  \texttt{1312.0634}

\bibitem{Jenkins:2013zja}
E.~E. Jenkins, A.~V. Manohar, M.~Trott, JHEP \textbf{10}, 087 (2013),
  \texttt{1308.2627}

\bibitem{Jenkins:2013wua}
E.~E. Jenkins, A.~V. Manohar, M.~Trott, JHEP \textbf{01}, 035 (2014),
  \texttt{1310.4838}

\bibitem{Alonso:2013hga}
R.~Alonso, E.~E. Jenkins, A.~V. Manohar, M.~Trott, JHEP \textbf{04}, 159 (2014),
  \texttt{1312.2014}

\bibitem{Pruna:2014asa}
G.~M. Pruna, A.~Signer, JHEP \textbf{10}, 014 (2014), \texttt{1408.3565}

\bibitem{Pruna:2015jhf}
G.~M. Pruna, A.~Signer, EPJ Web Conf. \textbf{118}, 01031 (2016),
  \texttt{1511.04421}

\bibitem{Crivellin:2016ebg}
A.~Crivellin, S.~Davidson, G.~M. Pruna, A.~Signer, \emph{{Complementarity in
  lepton-flavour violating muon decay experiments}}, in \emph{{18th
  International Workshop on Neutrino Factories and Future Neutrino Facilities
  Search (NuFact16) Quy Nhon, Vietnam, August 21-27, 2016}} (2016),
  \texttt{1611.03409}

\bibitem{Crivellin:2017rmk}
A.~Crivellin, S.~Davidson, G.~M. Pruna, A.~Signer, JHEP \textbf{05}, 117 (2017),
  \texttt{1702.03020}

\bibitem{Davidson:2016edt}
S.~Davidson, Eur. Phys. J. \textbf{C76}, 370 (2016), \texttt{1601.07166}

\bibitem{Shanker:1979ap}
O.~U. Shanker, Phys. Rev. \textbf{D20}, 1608 (1979)

\bibitem{Czarnecki:1998iz}
A.~Czarnecki, W.~J. Marciano, K.~Melnikov, AIP Conf. Proc. \textbf{435}, 409
  (1998), \texttt{hep-ph/9801218}

\bibitem{Kitano:2002mt}
R.~Kitano, M.~Koike, Y.~Okada, Phys. Rev. \textbf{D66}, 096002 (2002),
  [Erratum: Phys. Rev.D76,059902(2007)], \texttt{hep-ph/0203110}

\bibitem{Dohmen:1993mp}
C.~Dohmen et~al. (SINDRUM II), Phys. Lett. \textbf{B317}, 631 (1993)

\bibitem{Honecker:1996zf}
W.~Honecker et~al. (SINDRUM II), Phys. Rev. Lett. \textbf{76}, 200 (1996)

\bibitem{Cirigliano:2009bz}
V.~Cirigliano, R.~Kitano, Y.~Okada, P.~Tuzon, Phys. Rev. \textbf{D80}, 013002
  (2009), \texttt{0904.0957}

\bibitem{Hoferichter:2015dsa}
M.~Hoferichter, J.~Ruiz~de Elvira, B.~Kubis, U.~G. Meißner, Phys. Rev. Lett.
  \textbf{115}, 092301 (2015), \texttt{1506.04142}

\bibitem{Junnarkar:2013ac}
P.~Junnarkar, A.~Walker--Loud, Phys. Rev. \textbf{D87}, 114510 (2013),
  \texttt{1301.1114}

\bibitem{Bartolotta:2017mff}
A.~Bartolotta, M.~J. Ramsey--Musolf (2017), \texttt{1710.02129}

\bibitem{Suzuki:1987jf}
T.~Suzuki, D.~F. Measday, J.~P. Roalsvig, Phys. Rev. \textbf{C35}, 2212 (1987)

\bibitem{Misiak:1993es}
M.~Misiak, Phys. Lett. \textbf{B321}, 113 (1994), \texttt{hep-ph/9309236}

\bibitem{Ciuchini:1993fk}
M.~Ciuchini, E.~Franco, L.~Reina, L.~Silvestrini, Nucl. Phys. \textbf{B421}, 41
  (1994), \texttt{hep-ph/9311357}

\bibitem{tHooft:1972tcz}
G.~'t~Hooft, M.~J.~G. Veltman, Nucl. Phys. \textbf{B44}, 189 (1972)

\bibitem{Breitenlohner:1977hr}
P.~Breitenlohner, D.~Maison, Commun. Math. Phys. \textbf{52}, 11 (1977)

\bibitem{Koike:2010xr}
M.~Koike, Y.~Kuno, J.~Sato, M.~Yamanaka, Phys. Rev. Lett. \textbf{105}, 121601
  (2010), \texttt{1003.1578}

\bibitem{Uesaka:2016vfy}
Y.~Uesaka, Y.~Kuno, J.~Sato, T.~Sato, M.~Yamanaka, Phys. Rev. \textbf{D93},
  076006 (2016), \texttt{1603.01522}

\bibitem{Uesaka:2017yin}
Y.~Uesaka, Y.~Kuno, J.~Sato, T.~Sato, M.~Yamanaka (2017), \texttt{1711.08979}

\bibitem{Cirigliano:2017azj}
V.~Cirigliano, S.~Davidson, Y.~Kuno, Phys. Lett. \textbf{B771}, 242 (2017),
  \texttt{1703.02057}

\bibitem{Davidson:2017nrp}
S.~Davidson, Y.~Kuno, A.~Saporta (2017), \texttt{1710.06787}

\bibitem{Calame:2015fva}
C.~M. Carloni~Calame, M.~Passera, L.~Trentadue, G.~Venanzoni, Phys. Lett.
  \textbf{B746}, 325 (2015), \texttt{1504.02228}

\bibitem{Abbiendi:2016xup}
G.~Abbiendi et~al., Eur. Phys. J. \textbf{C77}, 139 (2017), \texttt{1609.08987}

\end{thebibliography}

\end{document}